150th Anniversary Celebration
Department of Civil Engineering
Bengal Engineering and Science University, Shibpur

International Conference

January 11-14, 2007

*Civil Engineering in the New Millennium:
Opportunities and Challenges*

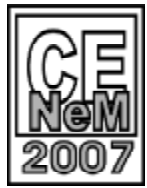

CENeM - 2007

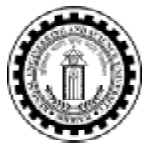

Organised by
Department of Civil Engineering
Bengal Engineering and Science University, Shibpur
Howrah-711103, West Bengal, India

# MARKOV CHAIN MODELLING FOR RELIABILITY ESTIMATION OF ENGINEERING SYSTEMS AT DIFFERENT SCALES – SOME CONSIDERATIONS

*K. Balaji Rao*

Scientist, Structural Engineering Research Centre
CSIR Campus, Taramani, Chennai 600 113, INDIA

**Abstract**

The concepts of probability, statistics and stochastic theory are being successfully used in structural engineering. Markov Chain modelling is a simple stochastic process model that has found its application in both describing stochastic evolution of system and in system reliability estimation. The recent developments in Markov Chain Monte Carlo and the possible integration of Bayesian theory within Markov Chain theory have enhanced its application possibilities. However, the application possibility can be furthered to range over wider scales of application (perhaps from nano- to macro-) by considering the developments in Physics (in particular Quantum Physics). This paper tries to present the results of quantum physics that would help in interpretation of transition probability matrix. However, care has to be taken in the choice of densities in computing the transition probability matrix. The paper is based on available literature, and the aim is only to make an attempt to show how Markov Chain can be used to model systems at various scales.

## INTRODUCTION

Determination of reliability of structural components/systems subjected to static loading or slowly varying loads and whose response evolves nonlinearly with load or time is the topic of current research. This requires that the probabilistic characteristics of response of the system be known at any stage of loading (or at any time instant considered). The probabilistic response in turn characterizes the probability of finding the system in different states at the instant considered. It is known that the probability density function of response, at different stages of loading or time instants, can be determined by having a suitable deterministic response evolution method in conjunction with Monte Carlo simulation technique. This can become computationally intensive/expensive, especially for large size problems. In such cases, there is a need to develop a suitable methodology for predicting the response evolution over the range of index space of interest. Needless to mention, the methodology should take into account the deterministic process along with the stochastic characteristics of evolution through proper statistical (viz. moments, joint moments) and memory (viz. one-step or in general m-step) properties. Use of Markov Chains (MCs) for describing the evolution of stochastic processes is well accepted [Benjamin and Cornell, 1970, Ang and Tang, 1984, Ditlevsen and Madsen, 1996, Bogdanoff and Kozin, 1985, Prakash Desayi and Balaji Rao, 1989, Balaji Rao and Appa Rao, 2004, Balaji Rao, 2005]. However, there are issues that need to be addressed when MC is used to describe the evolution of nonlinear response process. One of the important issues is to understand the degree of correlation among the states between two successive time (or loading) steps/stages and how this correlation propagates through evolution of the states at different time (or loading) steps/stages over the range of interest. Some of these issues have been recently addressed at SERC [Balaji Rao and Appa Rao, 1999, 2001, 2004, Balaji Rao et al, 2004a, 2004b, 2004c, 2005, Balaji Rao, 2005].

This paper makes an attempt to address the following theoretical/analytical issues regarding Markov Chain modelling for describing nonlinear response evolution of systems (under static loading conditions) and presents a methodology for estimation of reliability of classical statistical systems. The reliability estimation is addressed as a first passage problem. The aim is not to present any examples since they are available in many references (viz. [Balaji Rao and Appa Rao, 1999, 2001, 2004, Balaji Rao et al, 2004a, 2004b, 2004c, 2005, Rocha and Schueller, 2005, Balaji Rao, 2005]).

While application of MC for stochastic modelling is well accepted in engineering for systems which are described using classical statistical mechanics, its usefulness in modelling systems at various scales is still an issue receiving recent attention [Beckerman, 1998, Barabasi, 2005, Melnyk et al, 2006]. In this paper, this issue is addressed by defining a metric and through identification of isometries associated with space-time symmetries and the use of these concepts for reversible systems. First a metric is defined in the state space of a quantum mechanical system (micro-scale) and it is shown how this metric also represents the statistical distance in a statistical system (macro-scale). The next task is to identify the isometries associated with the space time symmetries that facilitates in identifying the isometries associated with evolution of time reversible systems.

## APPLICATION OF MARKOV CHAIN FOR CLASSICAL STATISTICAL MECHANICAL SYSTEMS

### Markov Chain – Preliminaries

The space-time evolution of most systems can be described by a MC representation. The time reversible systems whose stochastic evolution can be described by a homogeneous Markov chain (HMC) can be used for representation of linear elastic systems. It is also well accepted to approximate nonlinearly evolving systems by piece-wise linear approximation. Thus, theory of HMC can be also used for nonhomogeneous Markov chain (NHMC), but the transition probability matrix (TPM) needs to be evaluated for different time/loading stages. Keeping this in view, first the preliminaries of HMC are presented. Then, formulations related to determination of reliability of systems whose response evolves nonlinearly with index space are presented.

Markov Chain (MC) is a stochastic process model in which both the state space and the index space are discrete [Benjamin and Cornell, 1970]. The stochastic evolution of the system, modelled by homogeneous MC can be completely described by the Transition Probability Matrix (TPM), $P$. For a system with $N$ states, the transition probability matrix $P$ is

$$P = \begin{bmatrix} p_{11} & p_{12} & \cdots & p_{1N} \\ p_{21} & p_{22} & \cdots & p_{2N} \\ \vdots & \vdots & & \vdots \\ p_{N1} & p_{N2} & \cdots & p_{NN} \end{bmatrix} \qquad (1)$$

Typically, the element $p_{ij}$ represents the probability that the state of the system will be $j$ in the next step given that the state at present is $i$. If $X_n$ denotes the state of the system at time $t_n$, then

$$p_{ij} = P(X_{k+1} = j \mid X_k = i) \qquad (2)$$

$P$ is a one-step transition probability matrix the elements of which can be evaluated analytically once the deterministic model of the evolutionary phenomenon is available. The $n$-step TPM is given by

$$P^n = P \times P \times \ldots \times P \; (n\text{-times}) \qquad (3)$$

where $P = \{p_{ij}\}_{i \in 1,N; \, j \in 1,N}$; $P^n = \{p_{ij}(n)\}_{i \in 1,N; \, j \in 1,N}$; $p_{ij}(n)$ is the Probability{system state will by '$j$' after n-transitions, starting from state '$i$'}. That is,

$$p_{ij}(n) = P(X_{k+n} = j \mid X_k = i) \qquad (4)$$

As the states of the system are mutually exclusive and collectively exhaustive after each transition, the probabilities in each row add up to 1.0. For a system whose evolution is defined by a homogeneous MC, the state of the system at any future time can be determined using the one-step TPM, once the initial state is known. The long-run behaviour of the system described by TPM, $P$, is characterised by a vector $\pi = (\pi_1, \pi_2, \ldots, \pi_N)$. This vector also characterises the steady-state behaviour of the system. The elements of the vector are obtained by solving the following set of simultaneous equations:

$$\left. \begin{array}{l} \pi P = \pi \\ \sum_{i=1}^{N} \pi_i = 1 \end{array} \right\} \qquad (5)$$

In general, the following are the steps involved in modelling the stochastic behaviour of the system using MC:

- Identification and definition of states of the system
- Determination of elements of TPM (may be obtained either using the observation or using analytical model developed or both)
- Estimation of the quantities of interest using the TPM
- Interpretation of the results obtained

If the phenomenon considered is nonlinear/nonhomogeneous, a nonhomogeneous MC needs to be used for modelling the stochastic evolution. In such a case, the n-step TPM is given by

$$P^n = P_1 \times P_2 \times P_3 \times \cdots \times P_{n-1} \tag{6}$$

where $P_m$ ($m=1, \ldots, n-1$) represents the 1-step TPM estimated at the $m^{th}$ stage.

*Determination of elements of TPM*

A typical element of TPM of Markov Chain considered can be written as,

$$p_{ij}(Y_k, Y_{k+1}) = \frac{P\{X(Y_{k+1}) = j \cap X(Y_k) = i\}}{P\{X(Y_k) = i\}} \tag{7}$$

which gives the probability of the system being in state '$j$' at loading stage $Y_{k+1}$ given that the system was in state '$i$' at loading stage $Y_k$. To compute these probabilities, the information regarding jpdf of states at any two successive loading stages, ($Y_k$, $Y_{k+1}$), and also at loading stage, $Y_k$, should be known. Since it is difficult to generate this information from the test data, in the present investigation, it is assumed that states at successive loading stages follow bivariate normal distributions and at any loading stage, the state follows a normal distribution. It is also noted that when means, variances and correlation coefficient values between the states at two successive loading stages ($\rho_{N,N+1}$) are known, the maximum entropy distribution would be bivariate normal distribution [Kapur, 1993]. Knowing the jpdf and pdf, and using the above equation, the elements of TPM can be computed. A typical element of the conditional 1-step TPM is given by

$$p_{ij}(Y_k, Y_{k+1}) = \frac{\int_{x_{i-1}}^{x_i} \int_{x_{j-1}}^{x_j} f_{k,k+1}(x_k, x_{k+1}) dx_k dx_{k+1}}{\int_{x_{i-1}}^{x_i} f_k(x_k) dx_k} \tag{8}$$

where $f_{k,k+1}(x_k, x_{k+1})$ is the bivariate normal distribution given by

$$f_{k,k+1}(x_k, x_{k+1}) = \frac{1}{2\pi \sigma_k \sigma_{k+1} \sqrt{1 - \rho_{k,k+1}^2}} \exp\left\{-\frac{1}{2(1-\rho_{k,k+1}^2)}\left[\left(\frac{x_k - \mu_k}{\sigma_k}\right)^2 - 2\rho_{k,k+1}\left(\frac{x_k - \mu_k}{\sigma_k}\right)\left(\frac{x_{k+1} - \mu_{k+1}}{\sigma_{k+1}}\right) + \left(\frac{x_{k+1} - \mu_{k+1}}{\sigma_{k+1}}\right)^2\right]\right\} \tag{9}$$

$$-\infty \leq x_k \leq \infty; \; -\infty \leq x_{k+1} \leq \infty; \; -1 \leq \rho_{k,k+1} \leq 1;$$

and $f_k(x_k)$ is the univariate normal distribution, given by

$$f_k(x_k) = \frac{1}{\sqrt{2\pi}\sigma_k} \exp\left\{-\frac{1}{2}\left(\frac{x_k - \mu_k}{\sigma_k}\right)^2\right\}; \quad -\infty \leq \varepsilon_k \leq \infty \tag{10}$$

Equation 8 is general and can also be used to formulate the TPM for cases wherein the jpdf is other than bivariate normal.

*Determination of probabilistic behaviour and reliability of the system*

Using the formulations presented above, it is possible to compute the statistical properties of the response state of the system, at any stage of loading, by making use of following steps:

Divide the state space into mutually exclusive and collectively exhaustive event sets and compute the central value of each event set, namely, ($\Delta s_1, \Delta s_2, \Delta s_3, \ldots \Delta s_{m-1}, \Delta s_m$).

Compute unconditional probability vector of the states of the system at any stage of loading, that is, ($p_1, p_2, p_3, \ldots, p_m$) using Eq. 6.

Compute the mean, standard deviation and coefficient of variation of the response state of the system using,

$$\text{Mean response} = \langle S_{1,k} \rangle = \sum_{i=1}^{m} (\Delta s_i) p_i(Y_1, Y_k) \tag{11}$$

$$\text{Variance of response} = SD_{1,k}^2 = \sum_{i=1}^{m} (\Delta s_i)^2 p_i(Y_1, Y_k) - \langle S_{1,k} \rangle^2 \tag{12}$$

$$\text{cov} = \frac{SD_{1,k}}{\langle S_{1,k} \rangle} \tag{13}$$

It is noted that dividing state space into finite number of states depends on the nature of problem considered. If it is possible to determine the maximum possible (upper bound) response state of the system, the response statistics can be computed directly using Eqs. 11-13. However, if upper bound of the response state can not be computed, aggregation of response states should be done as follows. The allowable response, obtained based on specifications give in codes of practice, is considered as the starting point of the second aggregation state. Thus,

Aggregated State 1 = All states belonging to states with response less than the allowable response.

Aggregated State 2 = The state with starting value of the set is the allowable value.

When aggregation of response states is performed, it is possible to compute the probabilities of system in these two response states. It may be noted that the probability of the system in aggregated state 2 is the failure probability. Also, when aggregation is attempted, statistical properties (Eqs. 11-13) can be meaningful for aggregated state 1. These properties give an idea about the system performance in safe state. As can be expected the mean response of system in aggregated state 1 decreases with the increase of load.

The NHMC constructed to describe the response process of the system can be used to compute the reliability of the system, against the specified limit state at a stage of loading. The sum of probabilities of the system being in aggregated state 2 gives the failure probability of the system. Thus,

$$P_f(Y_k) = 1 - \sum_{i=1}^{k1} p_i(Y_k) \tag{14}$$

where $i = 1, 2, 3, \ldots, k1$ are states belonging to aggregated state 1, and $p_i(Y_k)$ are unconditional probabilities at applied loading of $Y_k$.

## A NATURAL METRIC[1] IN THE STATE SPACE

Let the following notation be defined:

S = Set of states of a quantum mechanical system[2]; $\Theta$ = Set of a real-valued observables;

$p(A, \alpha, E)$ = Function representing the probability that the measurement of observables $A \in \Theta$ on the state space $\alpha \in S$ given a result in the Borel set $E$ of the real numbers $\Re$. Thus, $p$ is defined on $\Theta \times S \times B$ (where $B$ is a family of Borel sets of $\Re$). $p$ has values in the closed interval [0,1] and is assumed to be such that $p(A, \alpha, \Re) = 1$, $p(A, \alpha, \phi) = 0$ (where $\phi$ is an empty set) and $p(A, \alpha, \cup E_b) = \Sigma p(A, \alpha, E_b)$ for every discrete family of disjoint Borel sets. Moreover, it is

---

[1] *Definition of a metric space [Apostol, 1996] : A metric is a non-empty set M of objects (called points) together with a function d from M x M to R (called the metric of the space) satisfying the following properties for all points x,y,z $\in M$ :*

*d(x,x) = 0*

*d(x,y) >0 if x $\neq$ y*

*d(x,y) = d(y,x)*

*d(x,y) = d(x,z) + d(z,y) (called the triangle inequality)*

*Sometimes a metric space is denoted by (M,d) to emphasize that both the set M and the metric d play a role in the definition of a metric space.*

[2] *It may be mentioned here that the states of the system need not be only pure states but also entangled states occurring due to decoherence. And the application of principles of quantum mechanics can be for estimation of reliability of systems with inheritance and can also be applied over large length scales (even upto meters) provided the energy levels of the system are closer enough to define density operator [Penrose, 2004].*

natural to assume that $\alpha = \beta$ whenever, $p(A, \alpha, E) = p(A, \beta, E) \ \forall A \in \Theta, E \in B$ and $A = B$ whenever $p(A, \alpha, E) = p(B, \alpha, E) \ \forall \alpha \in S, E \in B$.

Associating the following two point function $T_A(\alpha, \beta)$ with every observable $A$, the measure on $\Re$ can be defined as follows:

$$T_A = \left| \int_\Re d\sqrt{\alpha_A \beta_A} \right|^2 \tag{15}$$

where $\alpha_A$ is the probability measure on $\Re$ defined by

$$\alpha_A = \int_E d\alpha_A = p(A, \alpha, E) \tag{16}$$

$\beta_A$ is the probability measure on $\Re$ defined by

$$\beta_A = \int_E d\beta_A = p(A, \beta, E) \tag{17}$$

and, $\sqrt{\alpha_A \beta_A}$ denotes the measure on $\Re$ defined by

$$\int_E d\sqrt{\alpha_A \beta_A} = \int_E \sqrt{\frac{d\alpha_A}{d\sigma} \frac{d\beta_A}{d\sigma}} d\sigma \tag{18}$$

$\sigma$ being any finite measure on $\Re$ with respect to which $\alpha_A$ and $\beta_A$ are absolutely continuous. Next, by defining a two-point function $T(\alpha, \beta) = \inf_{\alpha \in \Theta} T_A(\alpha, \beta)$ and define the metric as follows

$$d(\alpha, \beta) = 2 \cos^{-1} \sqrt{T(\alpha, \beta)} \tag{19}$$

where $T(\alpha, \beta)$ in the above equations can be viewed as generalised transition probability in a quantum mechanical systems and, it actually coincides with transition probabilities between pure states. In full generality, however, there seems to be no reason why it should represent transition probability in any physical sense; rather, it should be regarded as quantitative evaluation of resemblance of the states $\alpha$ and $\beta$. This is particularly clear in statistical mechanics, where the states are represented by probability densities on the phase space $\Omega$, and the transition probability can be written down as

$$T(\alpha, \beta) = \left| \int_\Omega \sqrt{\rho_\alpha \rho_\beta} d\mu \right|^2 \tag{20}$$

where $\rho_\alpha$ and $\rho_\beta$ are probability density functions of $\alpha$ and $\beta$ and $\mu$ is the Liouville measure. Thus, $T(\alpha, \beta)$ in the above equation is a measure of or evaluation of overlap of $\rho_\alpha$ and $\rho_\beta$.

Since $T$ is symmetric, has values in the interval $[0,1]$, and is equal to 1 if and only if $\alpha = \beta$, the distance $d$ defined by Eq. 19 is also symmetric, not greater than $\pi$, and zero if and only if $\alpha \neq \beta$. It can be shown also that $d(\alpha, \beta)$ satisfies the triangle inequality so that $d(\alpha, \beta)$ endows S with a bounded metric (in this case $0 \leq d(\alpha, \beta) \leq \pi$). Like $T(\alpha, \beta)$, it provides a comparison between states: the smaller the distance, the greater the resemblance [Wootters, 1981].

In the special case where S represents the pure states of a quantum mechanical system, S is a complex projective space, and $d$ coincides with the geodesic distance determined from the Riemannian structure of S.

Symmetry forbids. Forbidding imposes order, but many different things that posses certain order may derive from the same symmetry. That is why physicists believe that the underlying symmetry, which forbids whole classes of occurrences at one stroke, is, in a sense, more fundamental than the individual occurrences themselves, and is worth discovering. Every symmetry leaves some thing unchanged. Such an unchanged quantity is called invariant by mathematicians and conserved quantity by physicists. This property is very general and is given by the Noether's theorem "for each symmetry, there is a corresponding conservation law" [Icke, 1988].

## ISOMETRIES CONNCETED WITH SPACE-TIME SYMMETRIES

If it is assumed that the system possesses a space-time symmetry group *G*, in the usual sense that there exists a class of physically equivalent reference frames related by *G* [Armstrong, 1988], there is a natural action of group on the space *G* of all realisable states and on the set Θ of all observables. In fact, if a state $\alpha$ is prepared in the reference frame *R* by means of certain experimental arrangement, and $g \in G$ transforms *R* into an equivalent frame *R'*, another state $\alpha' \equiv g\alpha$ can be prepared, in principle, by transferring the experimental arrangement from *R* to *R'*. Similarly, to an observable *A* measured in *R* by a certain apparatus there corresponds another observable $A' \equiv gA$ measured by the same apparatus transferred to *R'*.

The action of *G* on S determined by the maps

$$g : \quad S \to S \qquad \qquad (21)$$
$$\quad \alpha \to g\alpha$$

is isometric. The physical equivalence of reference frames entails the relation

$$p(A,\alpha,E) = p(gA, g\alpha, E) \qquad (22)$$

from the definitions (1) and (2) the following relation is obtained

$$T_A(\alpha,\beta) = T_{gA}(g\alpha, g\beta) \qquad (23)$$

and

$$T(\alpha,\beta) = \inf_{A \in \Theta} T_A(\alpha,\beta) = \inf_{gA \in g\Theta} T_{gA}(g\alpha, g\beta) = \inf_{A' \in \Theta} T_{A'}(g\alpha, g\beta) = T(g\alpha, g\beta) \qquad (24)$$

with bijective character of maps $g : \Theta \to \Theta$ taken into account. Thus, $T(\alpha,\beta) = T(g\alpha, g\beta)$, and from Eq. 19

$$d(\alpha, \beta) = d(g\alpha, g\beta), \quad g \in G \qquad (25)$$

As mentioned in the previous section, when S represents the pure states of a quantum-mechanical system, $T(\alpha,\beta)$ coincides with transition probability when the group is represented by unitary transformation of the underlying Hilbert space.

## ISOMETRIES CONNECTED WITH THE EVOLUTION OF REVERSIBLE SYSTEMS

Supposing that, from the point of view of a single reference frame, the generic state $\alpha$ of the system as the outcome of preparation ending at time $t_0$ and the generic observable *A* as a measurement process starting at time $t_0'$. Assuming, moreover, that for every $\tau \geq 0$ the environmental conditions can be controlled during the time interval ($t_0, t_0' \equiv t_0 + \tau$). Denoting by *m* the generic choice of a value of $\tau$ together with assignment of environmental conditions. Then, to each state $\alpha$ there corresponds another state $m\alpha$ which is the outcome of the preparation of $\alpha$ followed by the evolution under *m*, and to each observable *A* there corresponds another observable *mA* representing the same measurement process as *A*, but starting with delay $\tau$ after the system has evolved under *m*. It is natural that semi-group Ξ represents the set of all physically realisable movements m of the system, with actions on S and Θ subject to the identity $p(A,m\alpha,E) = p(mA,\alpha,E)$. It can be shown that $T(\alpha,\beta) \leq T(m\alpha, m\beta)$ for all $m \in \Xi$, or, equivalently,

$$d(\alpha, \beta) \geq d(m\alpha, m\beta) \qquad (26)$$

In the case of classical statistical mechanics the above relation is always equality, because the integral (Eq. 20) is invariant under the underlying canonical transformations associated with the movements of the system. In quantum mechanics the relation (Eq. 26) is again an equality; on account of unitary character of the evolution operators.

In greater generality, a sufficient condition in order that relation (Eq. 26) hold as an equality is the assumption that the movements Ξ constitute a group, rather than merely a semi-group. This can be interpreted as an assumption of reversibility, in the sense that to each regulation *m* of external conditions acting on the physical states one can associate another regulation $m^{-1}$ restoring the initial situation (with $m^{-1}$ independent of the state).

Thus, for reversible systems, the state space S possesses a group of isometries representing the movements of the physical system, and it is natural to assume that this group acts transitively on S: actually, the condition of transitivity can be regarded as the very definition of what is meant by assertion that distinct elements of S represent states of the same physical system.

## POSSIBILITIES FOR FUTURE

The following are essentially based on Wallace (2001).

**Description of states at equilibrium in classical statistical Mechanics:**

- The possible states of a classical statistical system are given by the points in some phase *P*.
- At any given time *t*, the specific system under consideration has a determinate state given by a specific point in *P* - though this point is assumed not to be exactly known.
- At time *t*, the probability that this determinate state is in a given region of *P* is given by some probability distribution over *P*
- The time – evolution of the system is deterministic (given by Hamilton's equations) and so knowing the probability distribution at one time tell us what it is at all other times.
- A system is said to be at equilibrium when the probability distribution does not vary in time.

On conceptual side, there is a problem of defining the probability distribution over phase space, though interpreted in a relative frequency terms. Because, the observed system is only one!.

With the above problem of non-existence of ensemble in a real world, Wallace (2001) suggests 'Quantum Interpretation of Statistical Probability (QISP)'.

**Quantum Interpretation of Statistical Probability (QISP).**

'Ignorance' probability in the sense of a probability distribution over a space of many possible states of a system, one of which is actual, has to be looked at critically in statistical mechanics. As such, the use of 'probability' density operator in statistical mechanics needs further examination. When a density operator, is used to describe a statistical system, it is to be understood as the determinate-though highly non-pure-'entanglement' density operator which describes that specific system.

The map is of from

$$p(\rho) \rightarrow \int D\rho \, p(\rho)\rho \tag{27}$$

Where *p(ρ)* is the given probability distribution over entanglement density operators *ρ* and the map in (27) is many-to-one. While map (27) is for a realistic quantum systems, to get a feel for '*ρ*' the form for an isolated quantum system is presented below.

$$\rho = \sum_i p(i) |i><i| \tag{28}$$

where *p(i)* is a given probability distribution *p(i)* over some (not necessarily orthogonal) states {|*i*>}.

The following six reasons for proposing the above conjecture were given in [Wallace, 2001]. Out of these six, the first three reasons are conceptual and the other three are more dynamical and probably more important.

1. In classical statistical mechanics, the main problem is under determination of probability distribution by the statistical facts. This problem would be automatically solved in QISP.

2. It would make the concept of 'ensemble' rather less problematic. By defining the density operator to be describing the system (single system under consideration) totally avoids the confusion of ensemble of classical statistical mechanics (which is more of a theoretical abstraction than a reality). In particular, concepts like entropy are defined, in CSM, to apply to an ensemble rather than an individual system, as in QISP. In quantum mechanics, if QISP holds, then it makes sense to describe a single system as being in a macrostate (i.e., described by an entanglement density operator), and we should be able to assign macrostate properties such as entropy to that single system. This may make at more coherent to describe a unique system as having ascertain probability distribution. This redescription of single systems has relevance for the reduction of thermodynamics to statistical mechanics.

3. If QISP holds, then the (highly problematic) probabilities of statistical mechanics are to a large extent removed from consideration, to be replaced with the probability intrinsic to quantum mechanics. However, this problem needs more research.

4. QISP allows us to construct 'transcendental' account of equilibrium-that is, a justification of the equilibrium state independent of any causal story as to how systems get into equilibrium in the first place- for quantum mechanics which in some way is similar to classical statistical mechanics. In the case of classical statistical mechanics the system equilibrium is decided in such way that the possible realizations of microstates are combined in such way that it is consistent with observed or to be modeled macrostate. Since we are considering equilibrium system behaviour we are talking about steady state modeling. The invariant quantity, assuming no dissipation, is energy. Hence, the candidate distributions proposed for microstates should be based on conservation of energy or should have energy as time invariant quantity. The microcanonical distribution hypothesized should satisfy the law of conservation of energy (it may be quickly recalled that the microcanonical distribution may be Boltzmann's distribution or equipartition distribution). In quantum mechanics also the concept of transcendental equilibrium is some what similar, except that in addition to above points (1)-(3), wherein we have density operator defined on states of quantum system (mostly entangled) are definite states of the system. Hence, some kind of eigen value analysis seems to help define the density operators on states of system. But all the studies from decoherence suggests that (in the absence of dissipation) the only density operators which are invariant under decoherence are projections (and sums of projections) onto eigenspaces of the conserved quantities. For a system with energy as the only conserved quantity, those invariant density operators are microcanonical operators and their sums.

5. Some of the important concepts, generally invoked in classical statistical mechanics, for describing the system in equilibrium is the concept of stationarity and much stronger property being ergodicity. Ergodicity is generally assumed to have mathematical simplicity/tractability and in engineering due to limitations imposed by experimentation (assuming that the process can be well approximated by a stationary process). The assumption of ergodicity is not required or it is natural to a quantum mechanical system since we neither have ensembles nor we have pdf evolving in time or constant defined over state space. We are handling a single system (dynamical) which is in equilibrium with environment (taken care of by decoherence of pure states of system).

6. If the plausibility of observation (4), dealing with equilibrium behaviour, is accepted, then the microcanonical density operator (interpreted as an entanglement density operator) is the only state of the system (at given energy) which is a valid equilibrium state-all other states evolve towards that state, so any probability distribution over any other states will not be an equilibrium distribution at all. In otherwords, QISP holds at equilibrium, because the dynamics of the system force it upon us.

## SUMMARY

In this paper, an attempt has been made to show how Markov Chain can be used to model systems at various scales. The emphasis was on how to address the formulation and interpretation of transition probability matrix using the concepts of classical statistical and quantum mechanics. It is to be mentioned that, at present, both classical statistical and quantum mechanics are to be applied depending upon the scales of phenomenon being modelled. To the author's knowledge, QISP seems to play an important role in future developments in experimental mechanics. Further studies in this direction are being carried out at SERC, Chennai.

## REFERENCES

Ang, A.H-S. and Tang, W.H. (1984), *Probability Concepts in Engineering Planning and Design: Vol.II - Decision, Risk and Reliability*, John Wiley and Sons.

Apostol, T.M. (1996), *Mathematical Analysis*, Narosa Publishing House.

Armstrong, M.A. (1988), *Groups and Symmetry*, Springer-Verlag.

Balaji Rao, K and Appa Rao, T.V.S.R. (1999), "A methodology for condition assessment of RC girder with limited inspection data", *The Bridge and Structural Engineer, Ing-IABSE*, 29(4), 13 – 26.

Balaji Rao, K and Appa Rao, T.V.S.R. (2001), "A methodology for reliability-based design of structural components subjected to fatigue loading", *Proceedings, National Symposium on Advances in Structural Dynamics and Design*, held at SERC, Madras, January 9 - 11, 2001, 401 - 407

Balaji Rao, K. and Appa Rao, T.V.S.R. (2004), "Stochastic modelling of crackwidth in reinforced concrete beams subjected to fatigue loading", *Engineering Structures*, 26(5), 665-673.

Balaji Rao, K., Anoop, M.B. and Lakshmanan, N. (2004a), "Modelling the evolutionary non-guassian processes using NHGMC", *Proceedings of the International Congress on Computational Mechanics and Simulation (ICCMS-2004)*, IIT, Kanpur, 9-12 December 2004, Vol. I: 182-189.

Balaji Rao, K., Anoop, M.B., Lakshmanan, N., Gopalakrishnan, S. and Appa Rao, T.V.S.R. (2004b), "Risk-based remaining life assessment of corrosion affected reinforced concrete structural members", *Journal of Structural Engineering*, 31(1), 51-64.

Balaji Rao, K., Anoop, M.B., Lakshmanan, N., Gopika Vinod, Saraf, R.K. and Kushwaha, H.S. (2004c), *A Methodology for Risk Informed In-Service Inspection for Safety Related Systems – Final Report*, Report No. SS-GAP01241-RR-04-3, Structural Engineering Research Centre, Chennai.

Balaji Rao, K. (2005), "Markov chain modelling of evolution of serviceability-related effects in flexural members", *Invited Presentation*, *Structural Engineering Convention (SEC-2005)*, Indian Institute of Science, Bangalore, 14-16 December, 2005 (in CD format).

Balaji Rao, K., Anoop, M.B. and Vaidyanathan, C.V. (2005), "Some studies on Markov Chain Monte Carlo", *International Conference on Computational and Experimental Engineering and Sciences (ICCES05)*, IIT, Madras, December 1-6, 2005 (in CD format).

Barabasi, A-L. (2005), "Taming complexity", *Nature* Physics, 1, 68 -70.

Beckerman, M. (1998), "Cooperativity and parallelism in mathematical models of brain function", *SIAM News*, 31(5), 1-6.

Benjamin, J.R., and Cornell, A.C. (1970), *Probability, Statistics and Decision for Civil Engineers*, McGraw-Hill.

Bogdanoff J.L., and Kozin F. (1985), *Probabilistic Models of Cumulative Damage*, John Wiley, New York.

Ditlevsen, O. and Madsen, H. O. (1996), *Structural Reliability Methods*, John Wiley & Sons, New York, 1996.

Icke, V. (1995), *The Force of Symmetry*, Cambridge University Press.

Kapur J.N. (1993), *Maximum Entropy Models in Science And Engineering*, Wiley Eastern Limited.

Melnyk, S.S., Usatenko, O.V. and Yampol'skii, V.A. (2006), "Memory functions of the additive Markoff chains: applications to complex dynamic systems", *Physica A*, 361, 405-415.

Penrose, R. (2004), *The Road to Reality: A Complete Guide to the Laws of the Universe*, Jonathan Cape.

Prakash Desayi and Balaji Rao, K. (1989), "Markov Chain Model for Cracking Behaviour of Reinforced Concrete Beams", *Journal of Structural Engineering, ASCE*, 115(9), 2129-2144.

Rocha, M.M. and Schueller, G.I. (2005), "Markoff chain modelling of NDI techniques", Fatigue and Fracture of Engineering Materials and Structurews, 28, 267-278.

Wallace, D. (2001), "Implications of quantum theory in the foundations of statistical Mechanics", *http://philsci-archive.pitt.edu* (last accessed on 02.10.2006)

Wootters, W.K. (1981), "Statistical distance and Hilbert space", *Physical Review D*, 23(2), 357 – 362.

**ACKNOWLEDGEMENT**

This paper is published with the kind permission of the Director, Structural Engineering Research Centre, Chennai. The author is grateful to his Ph.D. Thesis Guide, Prof. Prakash Desayi, and, to Dr T.V.S.R. Appa Rao and Dr N. Lakshmanan, former and present Directors of SERC, Chennai, for giving the encouragement. The author would like to thank Mr. M.B. Anoop for his help in preparing the manuscript.